\documentclass[%
preprint,
 amsmath,amssymb,
 aps,
floatfix,
]{revtex4-1}

\usepackage{graphicx}
\usepackage[caption=false]{subfig}
\usepackage{dcolumn}
\usepackage{bm}

\usepackage{pgfplots}
\pgfplotsset{compat=newest}
\pgfplotsset{plot coordinates/math parser=false}
\newlength\fheight
\newlength\fwidth
\usepackage[colorinlistoftodos]{todonotes}

\begin{document}

\preprint{version 2}

\title{Suspension and simple optical characterization of two-dimensional membranes}

\author{David B. Northeast}
 \email{david.northeast@queensu.ca}	
\author{Robert G. Knobel}%
 \email{knobel@queensu.ca}
\affiliation{%
 Department of Physics, Engineering Physics and Astronomy, Queen's University\\
}%


%


\begin{abstract} 
We report on a method for suspending two-dimensional crystal materials in an electronic circuit using an only photoresists and solvents. Graphene and NbSe${}_2$ are suspended tens of nanometers above metal electrodes with clamping diameters of several microns. The optical cavity formed from the membrane/air/metal structures enables a quick method to measure the number of layers and the gap separation using comparisons between the expected colour and optical microscope images. This characterization technique can be used with just an illuminated microscope with a digital camera which makes it adaptable to environments where other means of characterization are not possible, such as inside nitrogen glove boxes used in handling oxygen-sensitive materials.
\end{abstract}

\pacs{Valid PACS appear here}
\maketitle

\section{Motivation \label{sec_motive}}


The suspension of graphene and other two-dimensional (2D) materials away from contact with a substrate has enabled a wide range of fundamental discoveries and applications.  Even disregarding their remarkable electronic and optical properties, the fact that these atomically-thin materials can span large gaps allows them to be used as highly transparent supports for transition electron microscopy~\cite{Nair2010} or as vacuum-tight seals~\cite{Bunch2008}.  However, when embedded in an electric circuit, suspended 2D membranes can be used in many fundamental studies and promising applications.  The first demonstrations of the fractional quantum Hall effect in graphene~\cite{bolotin2009}~\cite{Du2009} were done using suspended graphene membranes, where the lack of substrate increases the electron mobility.  There are numerous applications for graphene and other thin membrane crystals in electromechanical and optomechanical systems, including as sensors of force, strain, or mass~\cite{WeberP02,Smith2013}, and as low-voltage mechanical switches ~\cite{NagaseM01,DragomanM01,SunJ01}.  Hexagonal boron nitride (HBN) and transition metal dichalcogenides (TMDs) broaden the range of electrical parameters available in the mechanical material, enabling access to semiconducting, insulating, metallic, and superconducting materials for optical or electrical coupling in the design of devices or experiments~\cite{RadisavljevicB01,SenguptaS1}. 

Fabricating and characterizing mechanical systems using suspended 2D crystal materials can be a difficult challenge.  The flexible membrane must be supported during processing, but this support must be removed to allow the suspension, all while keeping incompatible chemicals and processes away from the fragile membrane. Simply etching away silicon dioxide under graphene or TMD devices has been used in electromechanical resonators~\cite{SenguptaS1,ChenC01}, however these devices have limitations compared to locally gated devices.  Suspended structures using 2D materials and a recessed local electrode have been made by placing exfoliated membranes as the last fabrication step~\cite{SinghV01,WeberP01}, however more complicated circuits may require processing after the membrane has been placed.  After fabrication, characterization of the membrane's thickness and the suspended gap can be done using atomic force microscopy (AFM)~\cite{ShearerCJ01}, Raman spectroscopy~\cite{FerrariAC01} and scanning electron microscopy (SEM)~\cite{ChenC01}, however all these have the potential to damage the membrane~\cite{ElBanaMS01}.  Some promising 2D materials such as black phosphorus and NbSe$_2$ are air-sensitive~\cite{CaoY01}, so this characterization - if done in air - will cause even more damage.

In this work we introduce a novel fabrication process and rapid non-destructive characterization technique broadly applicable to suspended membranes of 2D materials.  Here a 2D membrane is integrated into an electronic circuit using resists for sacrificial layers and compatible chemicals for further processing.  The intended goal of this research is to use 2D membranes in experiments exploring the quantum regime of mechanical systems~\cite{WeberP01,SinghV01,SongX01,WillM01}, however the fabrication technique is more generally applicable.  In fabricating these devices, we require a quick, non-destructive method to characterize the thickness and suspended height of the membranes. We introduce a generalization of the commonly used technique  thin layers on a known thickness of SiO$_2$ over silicon~\cite{NovoselovK01,BlakeP01}, where thin film interference increases the graphene visibility. We extend this principle to measure both the membrane thickness and suspended gap without damaging electron beams, high intensity lasers or the risk of tearing due to an AFM tip. Here we use a colour camera with a white light source in a microscope operated in reflection mode and compare the resulting images with theoretical predictions based on the Fresnel equations.

\begin{figure*}[ht]
\includegraphics[width=0.9\textwidth]{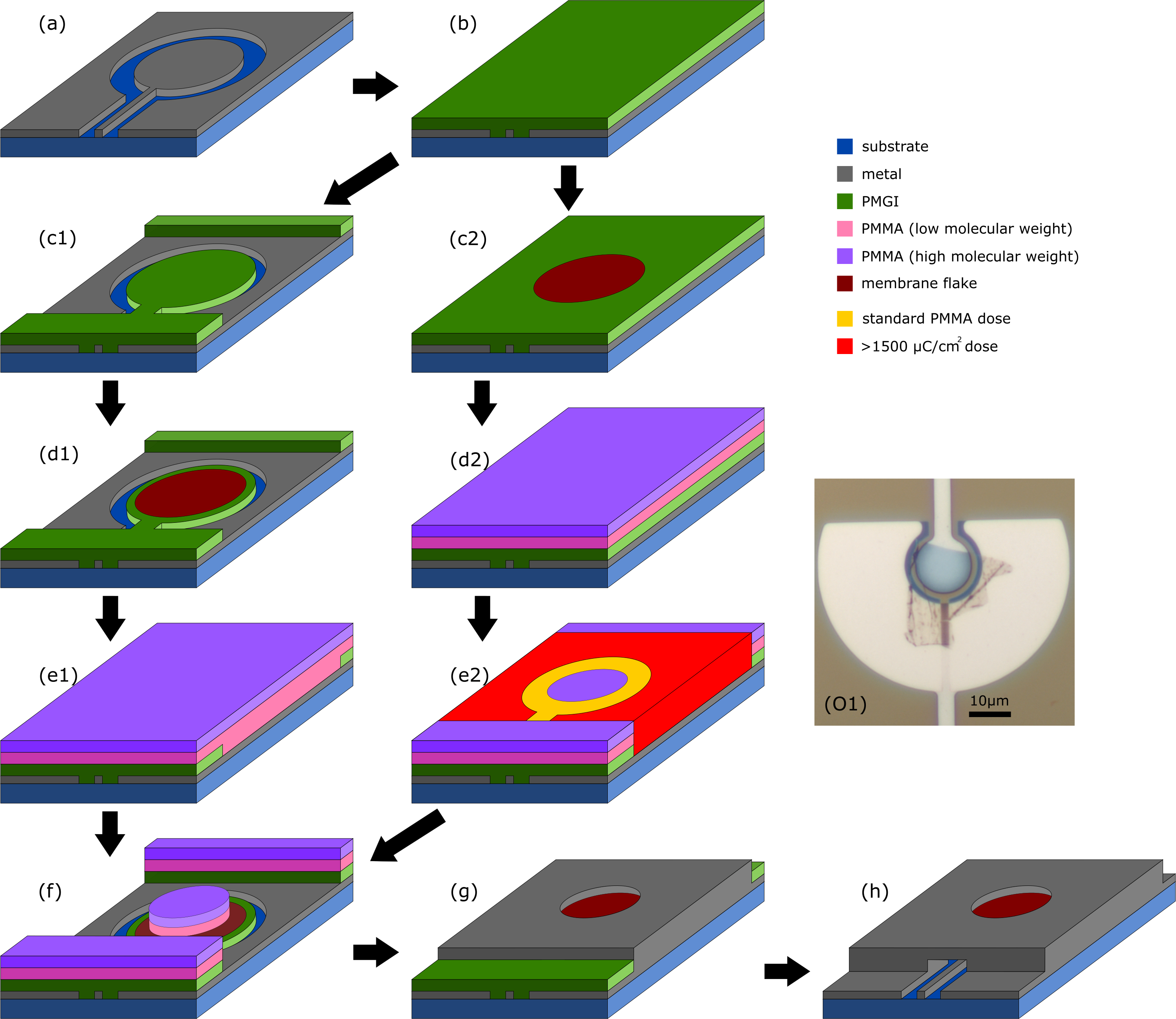}
\caption{The lithographic process to fabricate the suspended membrane structures starts with (a) a metallic structure on a substrate that is then (b) coated in PMGI. There are two variants to proceed. The first has the device patterned with e-beam lithography (c1) to open clamp windows to the metal. A 2D crystal membrane is placed over the resist-covered electrode (d1) and the chip is spin coated with (e1) two molecular weights of PMMA. This allows e-beam lithography to define clamping areas (f) for metal evaporation and lift off (g). The membrane is left freely suspended above the electrode by dissolving the PMGI in n-methyl-2-pyrrolidone and critical point drying (h). The second method has the membrane placed on the PMGI on step (c2), and the sample coated in two molecular weights of PMMA (d2). Step (e2) is a standard electron dose to pattern metal clamping areas for the membrane, and a high dose to pattern both the PMMA and PMGI following the process developed by Cui et al.~\cite{CuiB01}. Both processes are identical between steps (f)-(h). Image (O1) shows a stamped graphene membrane as in (d1).}
\label{fig_litho_process}
\end{figure*}

\section{Fabrication methods \label{sec_methods}}

We prepared samples using exfoliated membranes of graphene or 2H-NbSe${}_2$ taken from bulk crystals, thinned by several rounds of cleavage between two pieces of Nitto tape (SPV224). The thinned membranes are transferred onto polydimethylsiloxane (PDMS) sheets on a glass slide.  We identified and located even-coloured crystals that appear thin and unfolded using an optical microscope.   The glass slide is then placed membrane-side down in a modified mask aligner (Oriel) to position the flake accurately above a patterned silicon/silicon dioxide wafer coated with 120~nm of polymethylglutarimide (PMGI) electron-beam resist (Microchem).  We initiated the membrane transfer from the PDMS/glass slide to the PMGI-coated wafer by raising the wafer until it contacted the PDMS. Contact between the PDMS and wafer is observed through the mask aligner's microscope, though the glass and PDMS, appearing as a dark line advancing across the microscope field of view. We transfer the membrane from the PDMS to the PMGI-coated wafer by slowly lowering the wafer away after this contact.  While the PDMS is rigid for fast motions, it flows more like a liquid for slow motion, and the coated wafer retains the membrane as they separate. This transfer method is adapted from the method reported by Castellanos-Gomez {\em et al.}~\cite{CGomez02}. 

For our future experiments we want to have the 2D membrane form one plate of a vacuum-gap capacitor embedded in a resonant inductor-capacitor ($LC$) circuit.  Vibrations of the membrane change the capacitance, changing the resonant frequency, enabling a strong electro-mechanical coupling.  Similar systems using metallic thin films at cryogenic temperatures have proven to be a rich environment  for probing the quantum regime of mechanical systems~\cite{Teufel1}.  We are working to make lighter membranes using 2D materials where we suspend a graphene or NbSe$_2$ membrane about 100nm above an aluminum electrode[ref - your thesis in progress].  In the devices described here we place the graphene or NbSe$_2$ membrane above a circular aluminum electrode on the PMGI layer, and follow the procedure described below to make electrical connections and remove the PMGI, suspending the membrane.

\begin{figure}
	\includegraphics[width=0.8\columnwidth]{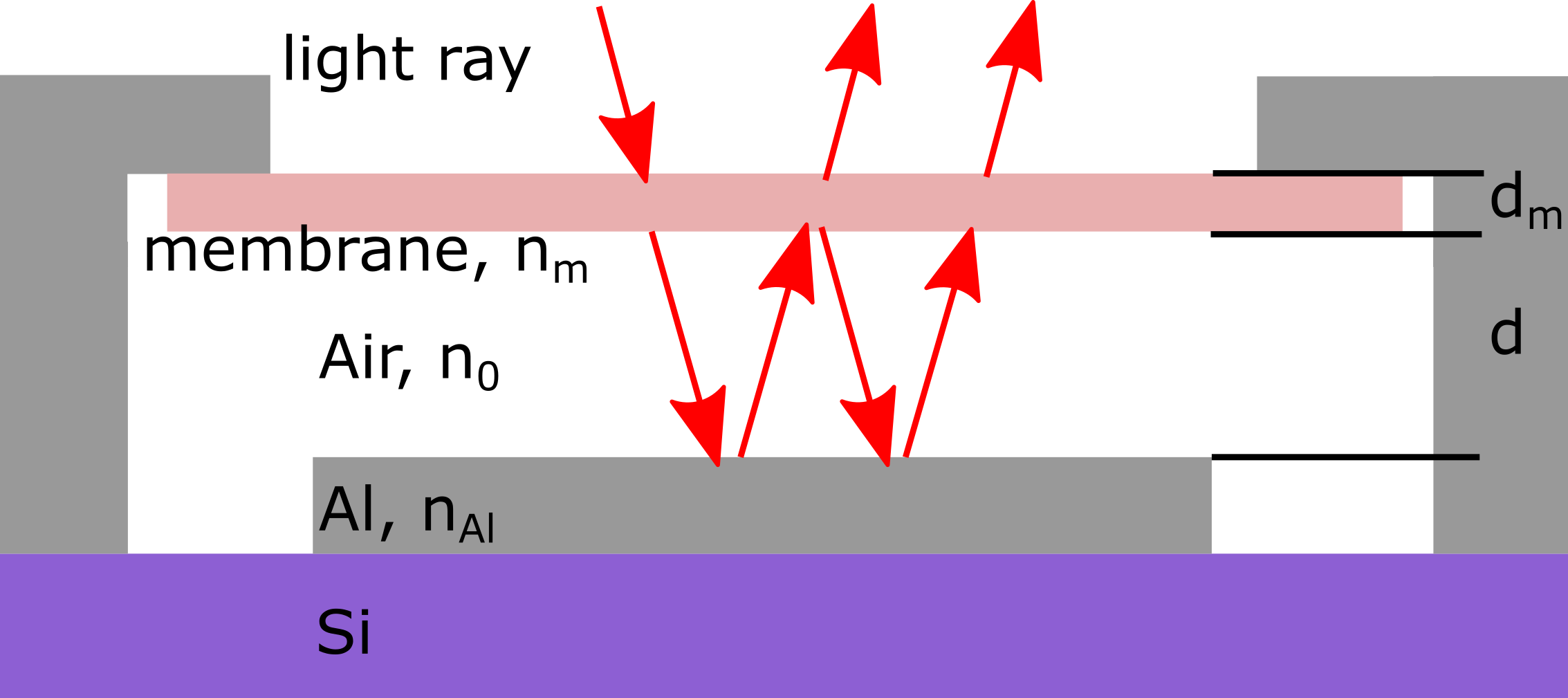}
	\caption{Cross-sectional schematic picture of a finished device and the characterization technique.  The separation $t_g$ is nominally set by the sacrificial resist thickness. Light will transmit and reflect at the interfaces of the different materials, causing interference which depends on the wavelength, thicknesses and indices of refraction.  }
	\label{fig_fab_finished}
\end{figure}

The lithography process is shown schematically in Fig.~\ref{fig_litho_process}. The process has two variations, and is based on selectively changing the solvent compatibility of polymethylglutarimide (PMGI) resist based on electron beam dose~\cite{CuiB01}. In both processes, a sacrificial layer (PMGI) coats the bottom layer of metal (Fig.~\ref{fig_litho_process}(b)).

The first method has the PMGI layer patterned with a standard dose ($\sim$300 $\mu$C/cm${}^2$) and developed in AZ developer. This patterning opens via windows into the bottom metal layer but leaves the electrode covered in resist (Fig.~\ref{fig_litho_process}(c1)). It is at this point that the membrane is stamped onto the chip (Fig.~\ref{fig_litho_process}(d1)), then a bilayer of polymethyl methacrylate (PMMA) is spin coated onto the sample (Fig.~\ref{fig_litho_process}(e1)). The bilayer consists of 950k molecular weight over 495k molecular weight PMMA that allows for an undercut profile and clean metal lift off (Fig.~\ref{fig_litho_process}(f)-(g)). Figure~\ref{fig_litho_process}(O1) is an optical image of a graphene membrane placed on PMGI, representing the step of Fig.~\ref{fig_litho_process}(c1).

The second method has the 2D crystal placed at step Fig.~\ref{fig_litho_process}(c2), then the two layers of PMMA are coated on top (Fig.~\ref{fig_litho_process}(d2)). The PMMA can be patterned with an electron beam dose of  $\sim$300 $\mu$C/cm${}^2$ (low dose) and developed in a solution of methyl isobutyl ketone (MIBK) and isopropyl alcohol (IPA) (1:3 ratio). At this dose, the PMGI is not soluble in MIBK/IPA and remains as a support for the membrane. If the applied dose is increased to $>1500 \mu$C/cm${}^2$ (high dose), the PMGI also becomes soluble in the MIBK/IPA mixture. This allows a low dose to define metal clamping structures over the membrane, then a high dose can be used to remove areas of developed PMGI where metal contacting vias are needed between top and bottom metal layers (Fig.~\ref{fig_litho_process}(e2)). The benefit of this method is that one step of electron-beam lithography can be used to create this structure, however the extra dose may dama

In both methods another layer of metal is deposited by electron-beam evaporation (Fig.~\ref{fig_litho_process}(f)-(g)) which connects the membrane electrically and provides mechanical support, allowing the sacrifical PMGI layer underneath to be removed.  We dissolve the PMGI in n-methyl-2-pyrrolidone (NMP), suspending the membrane from the top metal layer.  The sample is dried by slowly replacing NMP with acetone and then IPA and performing critical point drying in CO$_2$.

\begin{figure}
	\includegraphics[width=0.8\columnwidth]{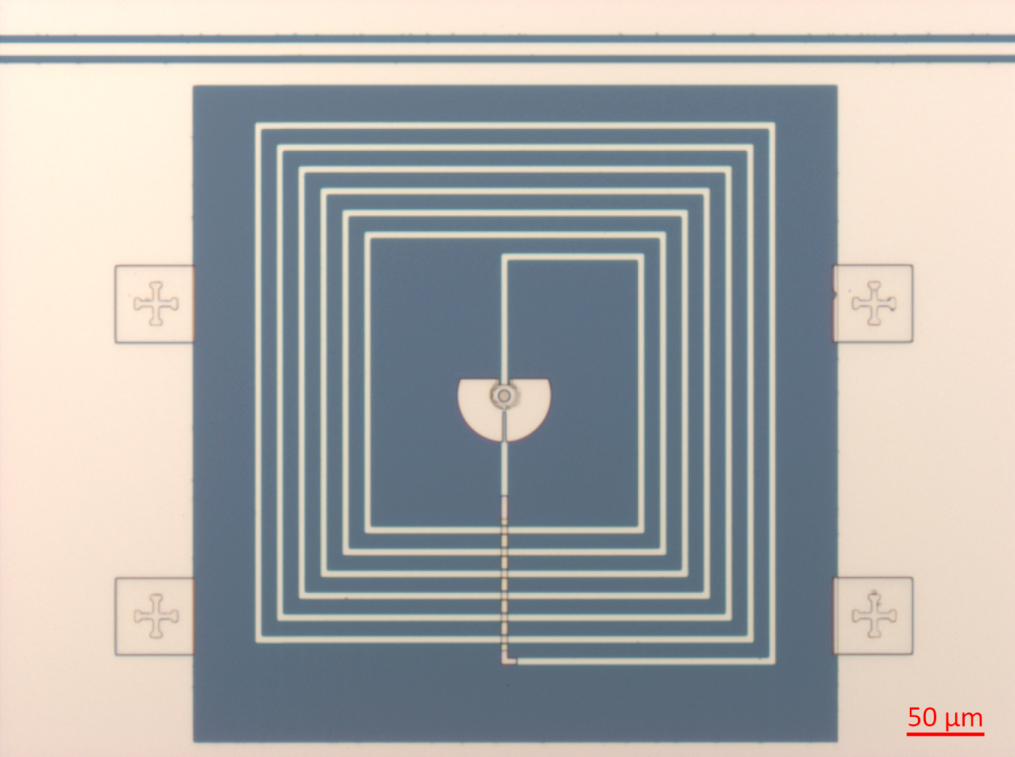}
	\caption{A completed device, showing a spiral inductor coupled to a transmission line (top).  A NbSe$_2$ membrane is placed over an electrode at the centre and air bridges connect the inner electrode to the outer loop.}
	\label{fig_air_bridges}
\end{figure}

This process also allows for the formation of air bridge structures to connect only where desired to the bottom metal layer. This can be seen in Fig.~\ref{fig_air_bridges}, where air bridges of aluminum help connect the inner portion of a spiral inductor to the outer portion.

\section{Interferometric colour analysis \label{sec_results}}

The colour of the membrane, as seen from top down on the sample, is a function of both the optical and geometric properties of the structure.  Measurement of this colour can give a rapid and non-destructive determination of both the thickness of the membrane $d_m$ (or number of layers $m = d_m/l$ where $l$ is the thickness of a single layer) as well as the gap it is suspended above the electrode $d$. 

Consider light of wavelength $\lambda$ (frequency $\omega$) illuminating the suspended membrane from above (figure \ref{fig_fab_finished}).  At the interface between air $n_0=1$ and the membrane with complex refractive index $\widetilde{n}_1 (\omega) = n_1 (\omega) + i \kappa_1 (\omega)$, a portion of the light will be reflected and a portion transmitted, given by the Fresnel equations~\cite{LevequeG01}
\begin{equation}
r_{0,1} = \frac{n_0 - n_1}{n_0+n_1}, \,\,\, t_{0,1}=1-r_{0,1},
\end{equation}
for normal incidence. This partial reflection and transmission occurs at the other surface of the membrane, as well as at the bottom aluminum electrode.  

We can use the transfer matrix method~\cite{KatsidisCC01,ByrnesSJ01} to determine the total reflection $r$ and transmission $t$ from the device. A wave traveling forward toward the substrate at normal incidence in a material of thickness $d_i$ with refractive index $\widetilde{n}_i$, has a wavevector given by $k_{z,i} = \frac{2 \pi \widetilde{n}_i}{\lambda_{0}}$. The total reflection ($r$) and transmission ($t$) from the stack can be related by
\begin{equation}
\begin{bmatrix}
1 \\
r
\end{bmatrix}
= \widetilde{M}
\begin{bmatrix}
t \\
0
\end{bmatrix},
\label{eq_total_r_and_t}
\end{equation}
with $\widetilde{M}$ given by
\begin{equation}
\widetilde{M} =\begin{bmatrix}
1 & r_{0,1} \\
r_{0,1} & 1
\end{bmatrix}M_1 M_2 \cdots M_{N-1}.
\label{eq_wave_matrix_all}
\end{equation}
For each of the $N$ layers of the stack, we have the $M_i$ matrix
\begin{equation}
M_i = \begin{bmatrix}
e^{-i d_i k_{z,i}} & 0\\
0 & 0
\end{bmatrix}
\begin{bmatrix}
1 & r_{i,i+1}\\
r_{i,i+1} & 1
\end{bmatrix}\frac{1}{d_{i,i+1}}.
\label{eq_wave_N}
\end{equation}
Over the visible range, the complex refractive indices of aluminum~\cite{RakicA01}, graphene~\cite{BlakeP01}, and NbSe${}_2$~\cite{BealAR01} have been well characterized.

The sensitivity of the reflectivity to thin layers of graphene over a Si/SiO$_2$ substrate has been used since its discovery  to rapidly identify the material\cite{NovoselovK01,BlakeP01}.  This effect is used here more generally.  When illuminated by white light, the varying reflectivity with wavelength gives the suspended membrane different colours depending on the membrane thickness and suspension height when observed with a microscope.  Measuring the reflectivity over all wavelengths would allow for both the membrane thickness and air gap to be determined.  We show here that even a colour camera without special filters can do the same.

We use a transfer matrix calculation program written in Python based on the \textit{tmm}~\cite{ByrnesSJ01} package to calculate reflectances over the optical spectrum. The Python package \textit{ColorPy}~\cite{python_colorpy} allows us to combine the reflectivity as a function of wavelength to simulate the apparent colour of the stack, and to compare both qualitatively and quantitatively to the fabricated devices.  

\begin{figure}
\includegraphics[width=0.75\columnwidth]{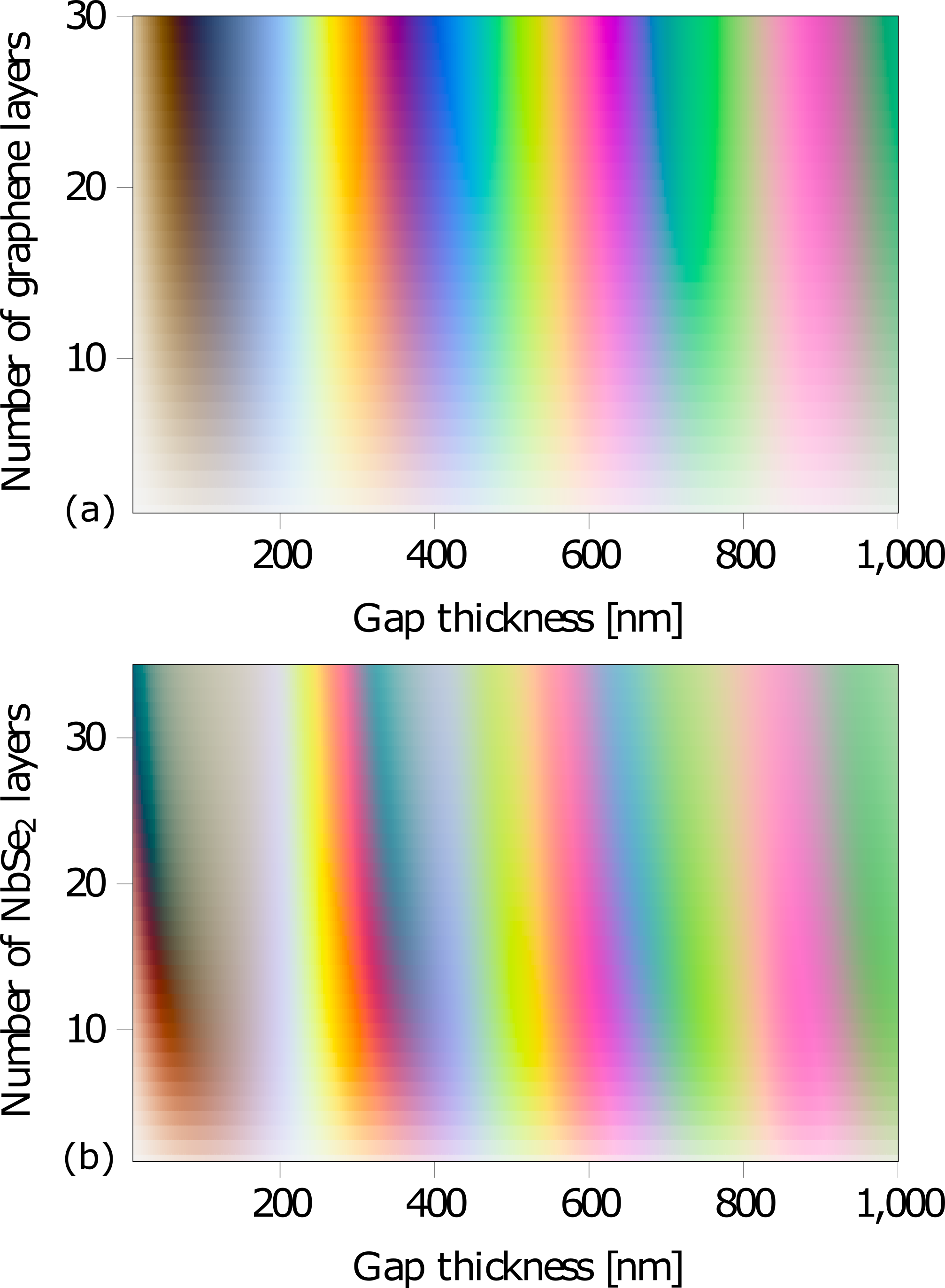}
\caption{Simulation of colour, as viewed from above, of a graphene/air/aluminum stack (a) and NbSe${}_2$/air/aluminum stack (b). The colour is a function of both the number of crystal layers $m$ and gap distance, $d$.}%
\label{fig_membrane_colour_d_vs_N}
\end{figure}

A stack consists of a semi-infinite air layer above the membrane, then an air-filled gap, then a semi-infinite aluminum layer.  We use the International Commission on Illumination's (CIE) D65 illuminant (white daylight) in the simulation. Fig.~\ref{fig_membrane_colour_d_vs_N}(a) and \ref{fig_membrane_colour_d_vs_N}(b) are plots showing the expected colours as the gap $d$ and number of layers $m$ are changed for both graphene and NbSe$_2$. There are periodic changes in expected colour as the gap distance are changed.  However the periodicity isn't perfect as the thickness or gap are changed, and distinctions can be made due to a change in the lightness of the colour. At larger $m$ (not shown in plots), the colour variations cease as the material reaches the colour of the bulk material.


We quantitatively measure the colours using images taken with a Carl Zeiss Axio Imager A1m microscope. Care is taken to colour balance the microscope camera against an unfocused image on white paper. In order to compare colours from the simulation to the image, we make use of the CIE colour difference function (CIEDE2000 implementation~\cite{SharmaG01}), typically noted as $\Delta E$. For $\Delta E$, the 24 bit (RGB) pixel values are converted from this colour space to one based on lightness (L), red-green opponent colours (a) and blue-yellow opponent colours (b), called the Lab colour space, which is required for $\Delta E$ comparisons.

\begin{figure*}
\includegraphics[width=\textwidth]{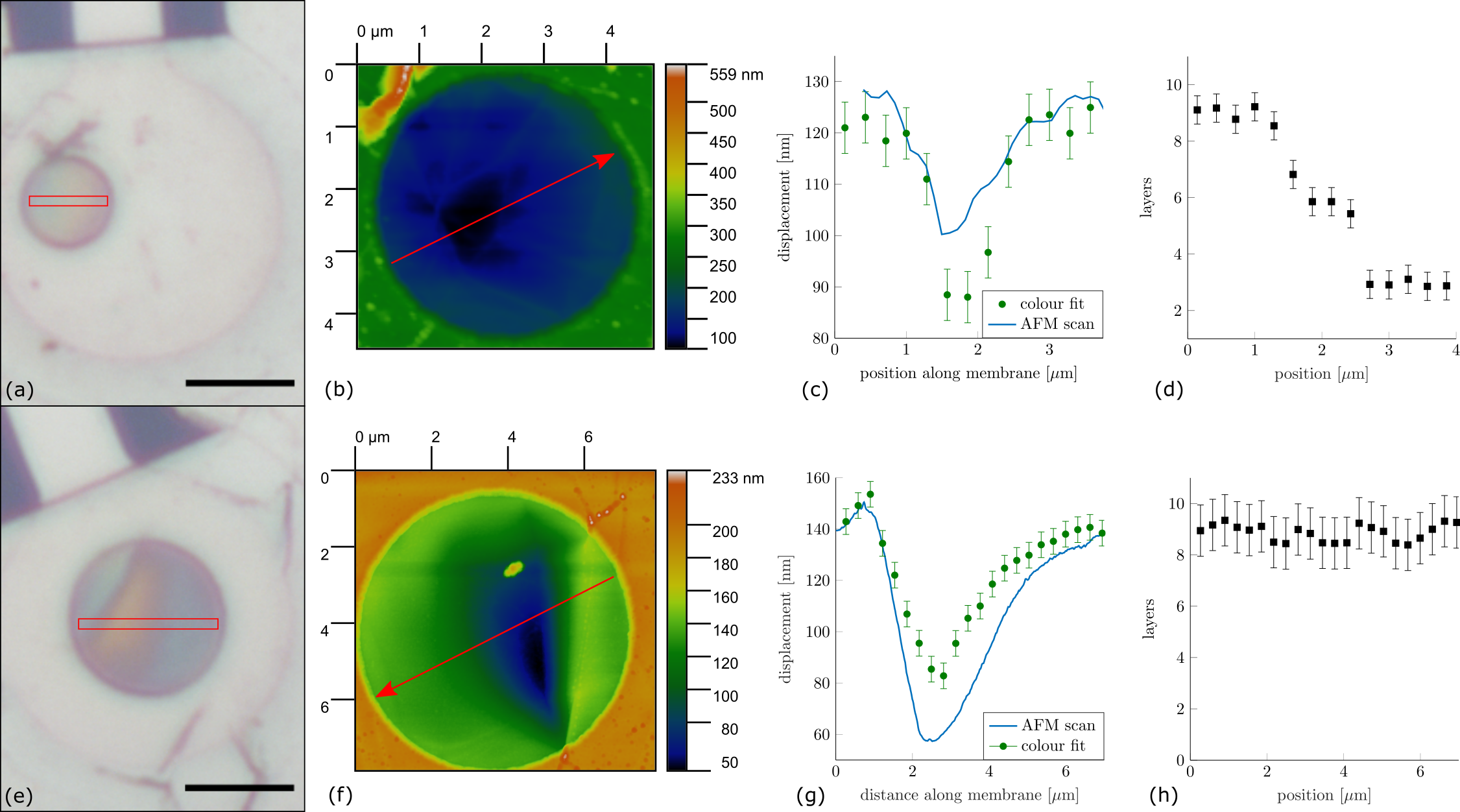}
\caption{Optical images of manufactured graphene devices (a,e) compared with atomic force microscopy (AFM) scans (b,f). Cross section plots of the number of membrane layers and gap distance are calculated using the colour comparison technique. Figure (c) and (g) plot the estimated gap separation against the measured AFM profile of the same graphene device (along the red arrow). The red rectangles in the circular membranes indicate the area, along the greater rectangle length, used in the cross sectional plots [with (c,d) corresponding to the cross section of (a), and (g,h) corresponding to the cross section of (e)]. The plots (d) and (h) show estimates of the number of membrane layers. Black scale bars in the optical images indicate a distance of 5 $\mu$m.}
\label{fig_optical_images}
\end{figure*}

\begin{figure*}
\includegraphics[width=\textwidth]{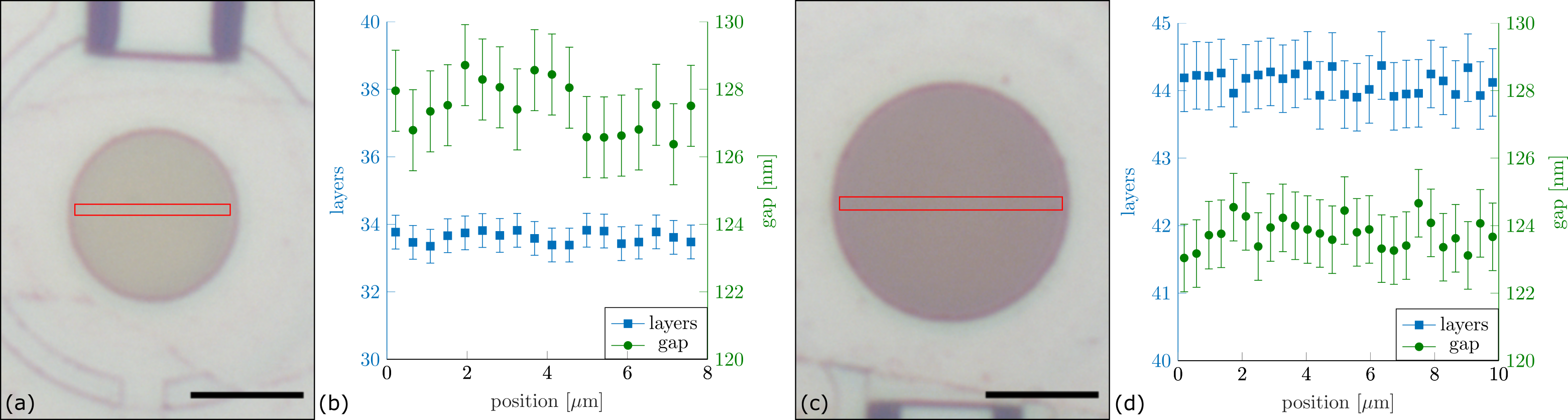}
\caption{Optical images of manufactured NbSe$_2$ devices and cross section plots of the number of membrane layers and gap distance. The red rectangles in the circular membranes indicate the areas used in the cross sectional plots [with (b) corresponding to the cross section of (a), and (d) the cross section of (c)]. The plots show estimates of the number of membrane layers and the gap thickness at positions along the greater length of the red rectangles. Black scale bars in the optical images indicate a distance of 5 $\mu$m. No AFM images were taken of the NbSe$_2$ membranes to avoid air or tip-induced damage.}
\label{fig_optical_images2}
\end{figure*}

Figures~\ref{fig_optical_images} and~\ref{fig_optical_images2} show some examples of devices made from graphene and NbSe${}_2$ respectively. In the graphene optical images we see large changes in colour due to wrinkles, sagging and a varying number of layers. The thicker NbSe$_2$ membranes are more rigid and show smaller variations in gap and thickness. Comparing the colours of the membranes in optical photos (Figures~\ref{fig_optical_images}(a) and (e) and~\ref{fig_optical_images2}(a) and (c)) with the simulation results in Fig.~\ref{fig_membrane_colour_d_vs_N}(a) and \ref{fig_membrane_colour_d_vs_N}(b), it is possible to fit the number of crystal layers and the gap separation by minimizing $\Delta E$. 

We take small sections of optical membrane photos to find the average and standard deviation of the RGB colour values.  Minimization of $\Delta E$ as a function of $m$ and $d$ is done through an optimization routine where the derivative or second derivative need not be specified. In this work, the Nelder-Mead method is used to minimize, and does not require the calculation of derivatives~\cite{scipy}.  Because the colour is non-monotonic, the minimization requires reasonable initial starting points for $d$ and $m$ so as not to converge on spurious points with similar colour.  We use the known thicknesses of the spin-coated resist and the evaporated metals to have good estimates of $d$ to start, and the transparency of the membrane gives a good starting point for $m$. Uncertainties in the fit values were determined statistically using the standard deviations of the measured colour as weights.

Breaking up an optical image into a series of smaller rectangular images, we fit to find of the number of membrane layers and the gap distance simultaneously across the suspended area. 
The optical colour measurement of gap and number of layers is performed along the diameter of the circle in the image (red rectangles in Fig.~\ref{fig_optical_images} and Fig.~\ref{fig_optical_images2}) which gives a line cut of gap separation. The number of layers for the graphene membranes is also estimated and plotted in Fig.~\ref{fig_optical_images}(d) and (h). The device in Fig.~\ref{fig_optical_images}(a) has some steps in the number of layers over the suspended area, with a variation in height of 35~nm across the diameter. The other graphene sample, device Fig.~\ref{fig_optical_images}(e) appears to be a constant $9.0 \pm 0.4$ membrane layers thick, with a noticeable 60~nm sag over the circle.

We validate these results by comparing to AFM images taken with an Asylum Research MFP-3D Origin+ AFM, shown in Fig.~\ref{fig_optical_images}(b) and (f).  The scans of the surface were in the attractive regime of tapping mode operation and the atomic force microscope (AFM) used Al-coated Si tips with a spring constant of 2~N/m.  The height recorded by AFM scan of the surface should equal $d+d_m$, and corresponds closely optically-determined to the gap separation $d$, as seen in figures~\ref{fig_optical_images}(c) and ~\ref{fig_optical_images}(g) which compare the line scans to the estimated colour gap fits along the diameter of the graphene.  The profiles constructed from the AFM scans and optical images follow the same trends, but there are discrepancies in the displacements determined from the two methods - though much of the difference can be attributed to averaging the optical signal over a larger area than the AFM measures.  Raman spectra of the graphene membranes (not shown) are consistent with membranes 5-10 layers thick, due to the shape of the 2D peak~\cite{FerrariAC01,MalardLM01}.


The same fitting is done with NbSe$_2$ membrane devices but without the comparison to AFM. NbSe$_2$ samples were thicker, with a constant gap distance as expected due to their even colour. The sample shown in Fig.~\ref{fig_optical_images2}(a) was determined to be $33.0 \pm 1.4$ layers with a gap of $126.0 \pm 2.2$~nm. The device seen in Fig.~\ref{fig_optical_images}(c) is $44.0 \pm 0.9$ layers with a gap of $123.1 \pm 3.0$~nm. The NbSe$_2$ samples (stored in a nitrogen glove box) were not scanned by the AFM as they are due to used in subsequent experiments and exposure to air while scanning the material may cause irreversible damage~\cite{CaoY01,ElBanaMS01}.




\section{Summary \label{sec_summary}}

We have developed a fabrication technique to create electromechanical devices using exfoliated crystal materials. The technique uses exfoliated membrane crystals that are aligned over an area coated in PMGI resist. Circular membrane sections with diameters of over 10~$\mu$m can be released and embedded in a larger circuit. The process uses a solvent release that avoids harsher techniques like hydrofluoric acid or reactive ion etching. 

We have generalized the well-known enhanced optical contrast of thin membranes due to thin films to study suspended 2D membranes.  Colours produced by optical interference can be measured using a simple digital camera, and can be used to measure both the thickness of the membrane and height of suspension, without the potential damage from scanned probes, high intensity lasers or electron microscopy.  These measurements are consistent with data from atomic force microscope images, and can be used for rapid and non-destructive characterization of suspended membranes.


\bibliography{thesis}

\begin{thebibliography}{34}%
\makeatletter
\providecommand \@ifxundefined [1]{%
 \@ifx{#1\undefined}
}%
\providecommand \@ifnum [1]{%
 \ifnum #1\expandafter \@firstoftwo
 \else \expandafter \@secondoftwo
 \fi
}%
\providecommand \@ifx [1]{%
 \ifx #1\expandafter \@firstoftwo
 \else \expandafter \@secondoftwo
 \fi
}%
\providecommand \natexlab [1]{#1}%
\providecommand \enquote  [1]{``#1''}%
\providecommand \bibnamefont  [1]{#1}%
\providecommand \bibfnamefont [1]{#1}%
\providecommand \citenamefont [1]{#1}%
\providecommand \href@noop [0]{\@secondoftwo}%
\providecommand \href [0]{\begingroup \@sanitize@url \@href}%
\providecommand \@href[1]{\@@startlink{#1}\@@href}%
\providecommand \@@href[1]{\endgroup#1\@@endlink}%
\providecommand \@sanitize@url [0]{\catcode `\\12\catcode `\$12\catcode
  `\&12\catcode `\#12\catcode `\^12\catcode `\_12\catcode `\%12\relax}%
\providecommand \@@startlink[1]{}%
\providecommand \@@endlink[0]{}%
\providecommand \url  [0]{\begingroup\@sanitize@url \@url }%
\providecommand \@url [1]{\endgroup\@href {#1}{\urlprefix }}%
\providecommand \urlprefix  [0]{URL }%
\providecommand \Eprint [0]{\href }%
\providecommand \doibase [0]{http://dx.doi.org/}%
\providecommand \selectlanguage [0]{\@gobble}%
\providecommand \bibinfo  [0]{\@secondoftwo}%
\providecommand \bibfield  [0]{\@secondoftwo}%
\providecommand \translation [1]{[#1]}%
\providecommand \BibitemOpen [0]{}%
\providecommand \bibitemStop [0]{}%
\providecommand \bibitemNoStop [0]{.\EOS\space}%
\providecommand \EOS [0]{\spacefactor3000\relax}%
\providecommand \BibitemShut  [1]{\csname bibitem#1\endcsname}%
\let\auto@bib@innerbib\@empty
\bibitem [{\citenamefont {Nair}\ \emph {et~al.}(2010)\citenamefont {Nair},
  \citenamefont {Blake}, \citenamefont {Blake}, \citenamefont {Zan},
  \citenamefont {Anissimova}, \citenamefont {Bangert}, \citenamefont
  {Golovanov}, \citenamefont {Morozov}, \citenamefont {Geim}, \citenamefont
  {Novoselov},\ and\ \citenamefont {Latychevskaia}}]{Nair2010}%
  \BibitemOpen
  \bibfield  {author} {\bibinfo {author} {\bibfnamefont {R.~R.}\ \bibnamefont
  {Nair}}, \bibinfo {author} {\bibfnamefont {P.}~\bibnamefont {Blake}},
  \bibinfo {author} {\bibfnamefont {J.~R.}\ \bibnamefont {Blake}}, \bibinfo
  {author} {\bibfnamefont {R.}~\bibnamefont {Zan}}, \bibinfo {author}
  {\bibfnamefont {S.}~\bibnamefont {Anissimova}}, \bibinfo {author}
  {\bibfnamefont {U.}~\bibnamefont {Bangert}}, \bibinfo {author} {\bibfnamefont
  {A.~P.}\ \bibnamefont {Golovanov}}, \bibinfo {author} {\bibfnamefont {S.~V.}\
  \bibnamefont {Morozov}}, \bibinfo {author} {\bibfnamefont {A.~K.}\
  \bibnamefont {Geim}}, \bibinfo {author} {\bibfnamefont {K.~S.}\ \bibnamefont
  {Novoselov}}, \ and\ \bibinfo {author} {\bibfnamefont {T.}~\bibnamefont
  {Latychevskaia}},\ }\href {\doibase 10.1063/1.3492845} {\bibfield  {journal}
  {\bibinfo  {journal} {Applied Physics Letters}\ }\textbf {\bibinfo {volume}
  {97}},\ \bibinfo {pages} {153102} (\bibinfo {year} {2010})},\ \Eprint
  {http://arxiv.org/abs/1010.4888} {arXiv:1010.4888} \BibitemShut {NoStop}%
\bibitem [{\citenamefont {Bunch}\ \emph {et~al.}(2008)\citenamefont {Bunch},
  \citenamefont {Verbridge}, \citenamefont {Alden}, \citenamefont {{Van Der
  Zande}}, \citenamefont {Parpia}, \citenamefont {Craighead},\ and\
  \citenamefont {McEuen}}]{Bunch2008}%
  \BibitemOpen
  \bibfield  {author} {\bibinfo {author} {\bibfnamefont {J.~S.}\ \bibnamefont
  {Bunch}}, \bibinfo {author} {\bibfnamefont {S.~S.}\ \bibnamefont
  {Verbridge}}, \bibinfo {author} {\bibfnamefont {J.~S.}\ \bibnamefont
  {Alden}}, \bibinfo {author} {\bibfnamefont {A.~M.}\ \bibnamefont {{Van Der
  Zande}}}, \bibinfo {author} {\bibfnamefont {J.~M.}\ \bibnamefont {Parpia}},
  \bibinfo {author} {\bibfnamefont {H.~G.}\ \bibnamefont {Craighead}}, \ and\
  \bibinfo {author} {\bibfnamefont {P.~L.}\ \bibnamefont {McEuen}},\ }\href
  {\doibase 10.1021/nl801457b} {\bibfield  {journal} {\bibinfo  {journal} {Nano
  letters}\ }\textbf {\bibinfo {volume} {8}},\ \bibinfo {pages} {2458}
  (\bibinfo {year} {2008})}\BibitemShut {NoStop}%
\bibitem [{\citenamefont {Bolotin}\ \emph {et~al.}(2009)\citenamefont
  {Bolotin}, \citenamefont {Ghahari}, \citenamefont {Shulman}, \citenamefont
  {Stormer},\ and\ \citenamefont {Kim}}]{bolotin2009}%
  \BibitemOpen
  \bibfield  {author} {\bibinfo {author} {\bibfnamefont {K.~I.}\ \bibnamefont
  {Bolotin}}, \bibinfo {author} {\bibfnamefont {F.}~\bibnamefont {Ghahari}},
  \bibinfo {author} {\bibfnamefont {M.~D.}\ \bibnamefont {Shulman}}, \bibinfo
  {author} {\bibfnamefont {H.~L.}\ \bibnamefont {Stormer}}, \ and\ \bibinfo
  {author} {\bibfnamefont {P.}~\bibnamefont {Kim}},\ }\href {\doibase
  10.1038/nature08582} {\bibfield  {journal} {\bibinfo  {journal} {Nature}\
  }\textbf {\bibinfo {volume} {462}},\ \bibinfo {pages} {196} (\bibinfo {year}
  {2009})}\BibitemShut {NoStop}%
\bibitem [{\citenamefont {Du}\ \emph {et~al.}(2009)\citenamefont {Du},
  \citenamefont {Skachko}, \citenamefont {Duerr}, \citenamefont {Luican},\ and\
  \citenamefont {Andrei}}]{Du2009}%
  \BibitemOpen
  \bibfield  {author} {\bibinfo {author} {\bibfnamefont {X.}~\bibnamefont
  {Du}}, \bibinfo {author} {\bibfnamefont {I.}~\bibnamefont {Skachko}},
  \bibinfo {author} {\bibfnamefont {F.}~\bibnamefont {Duerr}}, \bibinfo
  {author} {\bibfnamefont {A.}~\bibnamefont {Luican}}, \ and\ \bibinfo {author}
  {\bibfnamefont {E.~Y.}\ \bibnamefont {Andrei}},\ }\href {\doibase
  10.1038/nature08522} {\bibfield  {journal} {\bibinfo  {journal} {Nature}\
  }\textbf {\bibinfo {volume} {462}},\ \bibinfo {pages} {192} (\bibinfo {year}
  {2009})},\ \Eprint {http://arxiv.org/abs/0910.2532} {arXiv:0910.2532}
  \BibitemShut {NoStop}%
\bibitem [{\citenamefont {Weber}\ \emph {et~al.}(2016)\citenamefont {Weber},
  \citenamefont {{G{\"u}ttinger}}, \citenamefont {Noury}, \citenamefont
  {{Vergara-Cruz}},\ and\ \citenamefont {Bachtold}}]{WeberP02}%
  \BibitemOpen
  \bibfield  {author} {\bibinfo {author} {\bibfnamefont {P.}~\bibnamefont
  {Weber}}, \bibinfo {author} {\bibfnamefont {J.}~\bibnamefont
  {{G{\"u}ttinger}}}, \bibinfo {author} {\bibfnamefont {A.}~\bibnamefont
  {Noury}}, \bibinfo {author} {\bibfnamefont {J.}~\bibnamefont
  {{Vergara-Cruz}}}, \ and\ \bibinfo {author} {\bibfnamefont {A.}~\bibnamefont
  {Bachtold}},\ }\href {\doibase 10.1038/ncomms12496} {\bibfield  {journal}
  {\bibinfo  {journal} {Nature Communications}\ }\textbf {\bibinfo {volume}
  {7}},\ \bibinfo {pages} {12496} (\bibinfo {year} {2016})}\BibitemShut
  {NoStop}%
\bibitem [{\citenamefont {Smith}\ \emph {et~al.}(2013)\citenamefont {Smith},
  \citenamefont {Niklaus}, \citenamefont {Paussa}, \citenamefont {Vaziri},
  \citenamefont {Fischer}, \citenamefont {Sterner}, \citenamefont {Forsberg},
  \citenamefont {Delin}, \citenamefont {Esseni}, \citenamefont {Palestri},
  \citenamefont {{\"{O}}stling},\ and\ \citenamefont {Lemme}}]{Smith2013}%
  \BibitemOpen
  \bibfield  {author} {\bibinfo {author} {\bibfnamefont {A.~D.}\ \bibnamefont
  {Smith}}, \bibinfo {author} {\bibfnamefont {F.}~\bibnamefont {Niklaus}},
  \bibinfo {author} {\bibfnamefont {A.}~\bibnamefont {Paussa}}, \bibinfo
  {author} {\bibfnamefont {S.}~\bibnamefont {Vaziri}}, \bibinfo {author}
  {\bibfnamefont {A.~C.}\ \bibnamefont {Fischer}}, \bibinfo {author}
  {\bibfnamefont {M.}~\bibnamefont {Sterner}}, \bibinfo {author} {\bibfnamefont
  {F.}~\bibnamefont {Forsberg}}, \bibinfo {author} {\bibfnamefont
  {A.}~\bibnamefont {Delin}}, \bibinfo {author} {\bibfnamefont
  {D.}~\bibnamefont {Esseni}}, \bibinfo {author} {\bibfnamefont
  {P.}~\bibnamefont {Palestri}}, \bibinfo {author} {\bibfnamefont
  {M.}~\bibnamefont {{\"{O}}stling}}, \ and\ \bibinfo {author} {\bibfnamefont
  {M.~C.}\ \bibnamefont {Lemme}},\ }\href {\doibase 10.1021/nl401352k}
  {\bibfield  {journal} {\bibinfo  {journal} {Nano Letters}\ }\textbf {\bibinfo
  {volume} {13}},\ \bibinfo {pages} {3237} (\bibinfo {year} {2013})},\ \Eprint
  {http://arxiv.org/abs/1306.5876} {arXiv:1306.5876} \BibitemShut {NoStop}%
\bibitem [{\citenamefont {Nagase}\ \emph {et~al.}(2013)\citenamefont {Nagase},
  \citenamefont {Hibino}, \citenamefont {Kageshima},\ and\ \citenamefont
  {Yamaguchi}}]{NagaseM01}%
  \BibitemOpen
  \bibfield  {author} {\bibinfo {author} {\bibfnamefont {M.}~\bibnamefont
  {Nagase}}, \bibinfo {author} {\bibfnamefont {H.}~\bibnamefont {Hibino}},
  \bibinfo {author} {\bibfnamefont {H.}~\bibnamefont {Kageshima}}, \ and\
  \bibinfo {author} {\bibfnamefont {H.}~\bibnamefont {Yamaguchi}},\ }\href
  {http://stacks.iop.org/1882-0786/6/i=5/a=055101} {\bibfield  {journal}
  {\bibinfo  {journal} {Applied Physics Express}\ }\textbf {\bibinfo {volume}
  {6}},\ \bibinfo {pages} {055101} (\bibinfo {year} {2013})}\BibitemShut
  {NoStop}%
\bibitem [{\citenamefont {Dragoman}\ \emph {et~al.}(2009)\citenamefont
  {Dragoman}, \citenamefont {Dragoman}, \citenamefont {Coccetti}, \citenamefont
  {Plana},\ and\ \citenamefont {Muller}}]{DragomanM01}%
  \BibitemOpen
  \bibfield  {author} {\bibinfo {author} {\bibfnamefont {M.}~\bibnamefont
  {Dragoman}}, \bibinfo {author} {\bibfnamefont {D.}~\bibnamefont {Dragoman}},
  \bibinfo {author} {\bibfnamefont {F.}~\bibnamefont {Coccetti}}, \bibinfo
  {author} {\bibfnamefont {R.}~\bibnamefont {Plana}}, \ and\ \bibinfo {author}
  {\bibfnamefont {A.~A.}\ \bibnamefont {Muller}},\ }\href {\doibase
  10.1063/1.3080130} {\bibfield  {journal} {\bibinfo  {journal} {Journal of
  Applied Physics}\ }\textbf {\bibinfo {volume} {105}},\ \bibinfo {pages}
  {054309} (\bibinfo {year} {2009})}\BibitemShut {NoStop}%
\bibitem [{\citenamefont {Sun}\ \emph {et~al.}(2016)\citenamefont {Sun},
  \citenamefont {Muruganathan}, \citenamefont {Kanetake},\ and\ \citenamefont
  {Mizuta}}]{SunJ01}%
  \BibitemOpen
  \bibfield  {author} {\bibinfo {author} {\bibfnamefont {J.}~\bibnamefont
  {Sun}}, \bibinfo {author} {\bibfnamefont {M.}~\bibnamefont {Muruganathan}},
  \bibinfo {author} {\bibfnamefont {N.}~\bibnamefont {Kanetake}}, \ and\
  \bibinfo {author} {\bibfnamefont {H.}~\bibnamefont {Mizuta}},\ }\href
  {\doibase 10.3390/mi7070124} {\bibfield  {journal} {\bibinfo  {journal}
  {Micromachines}\ }\textbf {\bibinfo {volume} {7}},\ \bibinfo {pages} {124}
  (\bibinfo {year} {2016})}\BibitemShut {NoStop}%
\bibitem [{\citenamefont {Radisavljevic}\ \emph {et~al.}(2011)\citenamefont
  {Radisavljevic}, \citenamefont {Radenovic}, \citenamefont {Brivio},
  \citenamefont {Giacometti},\ and\ \citenamefont {Kis}}]{RadisavljevicB01}%
  \BibitemOpen
  \bibfield  {author} {\bibinfo {author} {\bibfnamefont {B.}~\bibnamefont
  {Radisavljevic}}, \bibinfo {author} {\bibfnamefont {A.}~\bibnamefont
  {Radenovic}}, \bibinfo {author} {\bibfnamefont {J.}~\bibnamefont {Brivio}},
  \bibinfo {author} {\bibfnamefont {V.}~\bibnamefont {Giacometti}}, \ and\
  \bibinfo {author} {\bibfnamefont {A.}~\bibnamefont {Kis}},\ }\href {\doibase
  10.1038/nnano.2010.279} {\bibfield  {journal} {\bibinfo  {journal} {Nature
  Nanotechnology}\ }\textbf {\bibinfo {volume} {6}},\ \bibinfo {pages} {147}
  (\bibinfo {year} {2011})}\BibitemShut {NoStop}%
\bibitem [{\citenamefont {{Sengupta}}\ \emph {et~al.}(2010)\citenamefont
  {{Sengupta}}, \citenamefont {{Solanki}}, \citenamefont {{Singh}},
  \citenamefont {{Dhara}},\ and\ \citenamefont {{Deshmukh}}}]{SenguptaS1}%
  \BibitemOpen
  \bibfield  {author} {\bibinfo {author} {\bibfnamefont {S.}~\bibnamefont
  {{Sengupta}}}, \bibinfo {author} {\bibfnamefont {H.~S.}\ \bibnamefont
  {{Solanki}}}, \bibinfo {author} {\bibfnamefont {V.}~\bibnamefont {{Singh}}},
  \bibinfo {author} {\bibfnamefont {S.}~\bibnamefont {{Dhara}}}, \ and\
  \bibinfo {author} {\bibfnamefont {M.~M.}\ \bibnamefont {{Deshmukh}}},\ }\href
  {\doibase 10.1103/PhysRevB.82.155432} {\bibfield  {journal} {\bibinfo
  {journal} {Physical Review B}\ }\textbf {\bibinfo {volume} {82}},\ \bibinfo
  {pages} {155432} (\bibinfo {year} {2010})}\BibitemShut {NoStop}%
\bibitem [{\citenamefont {Chen}\ \emph {et~al.}(2009)\citenamefont {Chen},
  \citenamefont {Rosenblatt}, \citenamefont {Bolotin}, \citenamefont {Kalb},
  \citenamefont {Kim}, \citenamefont {Kymissis}, \citenamefont {Stormer},
  \citenamefont {Heinz},\ and\ \citenamefont {Hone}}]{ChenC01}%
  \BibitemOpen
  \bibfield  {author} {\bibinfo {author} {\bibfnamefont {C.}~\bibnamefont
  {Chen}}, \bibinfo {author} {\bibfnamefont {S.}~\bibnamefont {Rosenblatt}},
  \bibinfo {author} {\bibfnamefont {K.~I.}\ \bibnamefont {Bolotin}}, \bibinfo
  {author} {\bibfnamefont {W.}~\bibnamefont {Kalb}}, \bibinfo {author}
  {\bibfnamefont {P.}~\bibnamefont {Kim}}, \bibinfo {author} {\bibfnamefont
  {I.}~\bibnamefont {Kymissis}}, \bibinfo {author} {\bibfnamefont {H.~L.}\
  \bibnamefont {Stormer}}, \bibinfo {author} {\bibfnamefont {T.~F.}\
  \bibnamefont {Heinz}}, \ and\ \bibinfo {author} {\bibfnamefont
  {J.}~\bibnamefont {Hone}},\ }\href {\doibase 10.1038/nnano.2009.267}
  {\bibfield  {journal} {\bibinfo  {journal} {Nature Nanotechnology}\ }\textbf
  {\bibinfo {volume} {4}},\ \bibinfo {pages} {861 } (\bibinfo {year}
  {2009})}\BibitemShut {NoStop}%
\bibitem [{\citenamefont {{Singh}}\ \emph {et~al.}(2014)\citenamefont
  {{Singh}}, \citenamefont {{Bosman}}, \citenamefont {Schneider}, \citenamefont
  {Blanter}, \citenamefont {{Castellanos-Gomez}},\ and\ \citenamefont
  {Steele}}]{SinghV01}%
  \BibitemOpen
  \bibfield  {author} {\bibinfo {author} {\bibfnamefont {V.}~\bibnamefont
  {{Singh}}}, \bibinfo {author} {\bibfnamefont {S.~J.}\ \bibnamefont
  {{Bosman}}}, \bibinfo {author} {\bibfnamefont {B.~H.}\ \bibnamefont
  {Schneider}}, \bibinfo {author} {\bibfnamefont {Y.~M.}\ \bibnamefont
  {Blanter}}, \bibinfo {author} {\bibfnamefont {A.}~\bibnamefont
  {{Castellanos-Gomez}}}, \ and\ \bibinfo {author} {\bibfnamefont {G.~A.}\
  \bibnamefont {Steele}},\ }\href {\doibase 10.1038/NNANO.2014.168} {\bibfield
  {journal} {\bibinfo  {journal} {Nature Nanotechnology}\ }\textbf {\bibinfo
  {volume} {9}},\ \bibinfo {pages} {820} (\bibinfo {year} {2014})}\BibitemShut
  {NoStop}%
\bibitem [{\citenamefont {Weber}\ \emph {et~al.}(2014)\citenamefont {Weber},
  \citenamefont {{G{\"u}ttinger}}, \citenamefont {Tsioutsios}, \citenamefont
  {Chang},\ and\ \citenamefont {Bachtold}}]{WeberP01}%
  \BibitemOpen
  \bibfield  {author} {\bibinfo {author} {\bibfnamefont {P.}~\bibnamefont
  {Weber}}, \bibinfo {author} {\bibfnamefont {J.}~\bibnamefont
  {{G{\"u}ttinger}}}, \bibinfo {author} {\bibfnamefont {I.}~\bibnamefont
  {Tsioutsios}}, \bibinfo {author} {\bibfnamefont {D.~E.}\ \bibnamefont
  {Chang}}, \ and\ \bibinfo {author} {\bibfnamefont {A.}~\bibnamefont
  {Bachtold}},\ }\href {\doibase 10.1021/nl500879k} {\bibfield  {journal}
  {\bibinfo  {journal} {Nano Letters}\ }\textbf {\bibinfo {volume} {14}},\
  \bibinfo {pages} {2854} (\bibinfo {year} {2014})}\BibitemShut {NoStop}%
\bibitem [{\citenamefont {Shearer}\ \emph {et~al.}(2016)\citenamefont
  {Shearer}, \citenamefont {Slattery}, \citenamefont {Stapleton}, \citenamefont
  {Shapter},\ and\ \citenamefont {Gibson}}]{ShearerCJ01}%
  \BibitemOpen
  \bibfield  {author} {\bibinfo {author} {\bibfnamefont {C.~J.}\ \bibnamefont
  {Shearer}}, \bibinfo {author} {\bibfnamefont {A.~D.}\ \bibnamefont
  {Slattery}}, \bibinfo {author} {\bibfnamefont {A.~J.}\ \bibnamefont
  {Stapleton}}, \bibinfo {author} {\bibfnamefont {J.~G.}\ \bibnamefont
  {Shapter}}, \ and\ \bibinfo {author} {\bibfnamefont {C.~T.}\ \bibnamefont
  {Gibson}},\ }\href {http://stacks.iop.org/0957-4484/27/i=12/a=125704}
  {\bibfield  {journal} {\bibinfo  {journal} {Nanotechnology}\ }\textbf
  {\bibinfo {volume} {27}},\ \bibinfo {pages} {125704} (\bibinfo {year}
  {2016})}\BibitemShut {NoStop}%
\bibitem [{\citenamefont {Ferrari}\ \emph {et~al.}(2006)\citenamefont
  {Ferrari}, \citenamefont {Meyer}, \citenamefont {Scardaci}, \citenamefont
  {Casiraghi}, \citenamefont {Lazzeri}, \citenamefont {Mauri}, \citenamefont
  {Piscanec}, \citenamefont {Jiang}, \citenamefont {Novoselov}, \citenamefont
  {Roth},\ and\ \citenamefont {Geim}}]{FerrariAC01}%
  \BibitemOpen
  \bibfield  {author} {\bibinfo {author} {\bibfnamefont {A.~C.}\ \bibnamefont
  {Ferrari}}, \bibinfo {author} {\bibfnamefont {J.~C.}\ \bibnamefont {Meyer}},
  \bibinfo {author} {\bibfnamefont {V.}~\bibnamefont {Scardaci}}, \bibinfo
  {author} {\bibfnamefont {C.}~\bibnamefont {Casiraghi}}, \bibinfo {author}
  {\bibfnamefont {M.}~\bibnamefont {Lazzeri}}, \bibinfo {author} {\bibfnamefont
  {F.}~\bibnamefont {Mauri}}, \bibinfo {author} {\bibfnamefont
  {S.}~\bibnamefont {Piscanec}}, \bibinfo {author} {\bibfnamefont
  {D.}~\bibnamefont {Jiang}}, \bibinfo {author} {\bibfnamefont {K.~S.}\
  \bibnamefont {Novoselov}}, \bibinfo {author} {\bibfnamefont {S.}~\bibnamefont
  {Roth}}, \ and\ \bibinfo {author} {\bibfnamefont {A.~K.}\ \bibnamefont
  {Geim}},\ }\href {\doibase 10.1103/PhysRevLett.97.187401} {\bibfield
  {journal} {\bibinfo  {journal} {Phys. Rev. Lett.}\ }\textbf {\bibinfo
  {volume} {97}},\ \bibinfo {pages} {187401} (\bibinfo {year}
  {2006})}\BibitemShut {NoStop}%
\bibitem [{\citenamefont {{El-Bana}}\ \emph {et~al.}(2013)\citenamefont
  {{El-Bana}}, \citenamefont {{Wolverson}}, \citenamefont {{Russo}},
  \citenamefont {{Balakrishnan}}, \citenamefont {{Paul}},\ and\ \citenamefont
  {{Bending}}}]{ElBanaMS01}%
  \BibitemOpen
  \bibfield  {author} {\bibinfo {author} {\bibfnamefont {M.~S.}\ \bibnamefont
  {{El-Bana}}}, \bibinfo {author} {\bibfnamefont {D.}~\bibnamefont
  {{Wolverson}}}, \bibinfo {author} {\bibfnamefont {S.}~\bibnamefont
  {{Russo}}}, \bibinfo {author} {\bibfnamefont {G.}~\bibnamefont
  {{Balakrishnan}}}, \bibinfo {author} {\bibfnamefont {D.~M.}\ \bibnamefont
  {{Paul}}}, \ and\ \bibinfo {author} {\bibfnamefont {S.~J.}\ \bibnamefont
  {{Bending}}},\ }\href {\doibase 10.1088/0953-2048/26/12/125020} {\bibfield
  {journal} {\bibinfo  {journal} {Superconductor Science and Technology}\
  }\textbf {\bibinfo {volume} {26}},\ \bibinfo {pages} {125020} (\bibinfo
  {year} {2013})}\BibitemShut {NoStop}%
\bibitem [{\citenamefont {Cao}\ \emph {et~al.}(2015)\citenamefont {Cao},
  \citenamefont {Mishchenko}, \citenamefont {Yu}, \citenamefont {Khestanova},
  \citenamefont {Rooney}, \citenamefont {Prestat}, \citenamefont {Kretinin},
  \citenamefont {Blake}, \citenamefont {Shalom}, \citenamefont {Woods},
  \citenamefont {Chapman}, \citenamefont {Balakrishnan}, \citenamefont
  {Grigorieva}, \citenamefont {Novoselov}, \citenamefont {Piot}, \citenamefont
  {Potemski}, \citenamefont {Watanabe}, \citenamefont {Taniguchi},
  \citenamefont {Haigh}, \citenamefont {Geim},\ and\ \citenamefont
  {Gorbachev}}]{CaoY01}%
  \BibitemOpen
  \bibfield  {author} {\bibinfo {author} {\bibfnamefont {Y.}~\bibnamefont
  {Cao}}, \bibinfo {author} {\bibfnamefont {A.}~\bibnamefont {Mishchenko}},
  \bibinfo {author} {\bibfnamefont {G.~L.}\ \bibnamefont {Yu}}, \bibinfo
  {author} {\bibfnamefont {E.}~\bibnamefont {Khestanova}}, \bibinfo {author}
  {\bibfnamefont {A.~P.}\ \bibnamefont {Rooney}}, \bibinfo {author}
  {\bibfnamefont {E.}~\bibnamefont {Prestat}}, \bibinfo {author} {\bibfnamefont
  {A.~V.}\ \bibnamefont {Kretinin}}, \bibinfo {author} {\bibfnamefont
  {P.}~\bibnamefont {Blake}}, \bibinfo {author} {\bibfnamefont {M.~B.}\
  \bibnamefont {Shalom}}, \bibinfo {author} {\bibfnamefont {C.}~\bibnamefont
  {Woods}}, \bibinfo {author} {\bibfnamefont {J.}~\bibnamefont {Chapman}},
  \bibinfo {author} {\bibfnamefont {G.}~\bibnamefont {Balakrishnan}}, \bibinfo
  {author} {\bibfnamefont {I.~V.}\ \bibnamefont {Grigorieva}}, \bibinfo
  {author} {\bibfnamefont {K.~S.}\ \bibnamefont {Novoselov}}, \bibinfo {author}
  {\bibfnamefont {B.~A.}\ \bibnamefont {Piot}}, \bibinfo {author}
  {\bibfnamefont {M.}~\bibnamefont {Potemski}}, \bibinfo {author}
  {\bibfnamefont {K.}~\bibnamefont {Watanabe}}, \bibinfo {author}
  {\bibfnamefont {T.}~\bibnamefont {Taniguchi}}, \bibinfo {author}
  {\bibfnamefont {S.~J.}\ \bibnamefont {Haigh}}, \bibinfo {author}
  {\bibfnamefont {A.~K.}\ \bibnamefont {Geim}}, \ and\ \bibinfo {author}
  {\bibfnamefont {R.~V.}\ \bibnamefont {Gorbachev}},\ }\href {\doibase
  10.1021/acs.nanolett.5b00648} {\bibfield  {journal} {\bibinfo  {journal}
  {Nano Letters}\ }\textbf {\bibinfo {volume} {15}},\ \bibinfo {pages} {4914}
  (\bibinfo {year} {2015})}\BibitemShut {NoStop}%
\bibitem [{\citenamefont {Song}\ \emph {et~al.}(2014)\citenamefont {Song},
  \citenamefont {Oksanen}, \citenamefont {Li}, \citenamefont {Hakonen},\ and\
  \citenamefont {{Sillanp{\"a}{\"a}}}}]{SongX01}%
  \BibitemOpen
  \bibfield  {author} {\bibinfo {author} {\bibfnamefont {X.}~\bibnamefont
  {Song}}, \bibinfo {author} {\bibfnamefont {M.}~\bibnamefont {Oksanen}},
  \bibinfo {author} {\bibfnamefont {J.}~\bibnamefont {Li}}, \bibinfo {author}
  {\bibfnamefont {P.~J.}\ \bibnamefont {Hakonen}}, \ and\ \bibinfo {author}
  {\bibfnamefont {M.~A.}\ \bibnamefont {{Sillanp{\"a}{\"a}}}},\ }\href
  {\doibase 10.1103/PhysRevLett.113.027404} {\bibfield  {journal} {\bibinfo
  {journal} {Physical Review Letters}\ }\textbf {\bibinfo {volume} {113}},\
  \bibinfo {pages} {027404} (\bibinfo {year} {2014})}\BibitemShut {NoStop}%
\bibitem [{\citenamefont {Will}\ \emph {et~al.}(2017)\citenamefont {Will},
  \citenamefont {Hamer}, \citenamefont {M{\"{u}}ller}, \citenamefont {Noury},
  \citenamefont {Weber}, \citenamefont {Bachtold}, \citenamefont {Gorbachev},
  \citenamefont {Stampfer},\ and\ \citenamefont {G{\"{u}}ttinger}}]{WillM01}%
  \BibitemOpen
  \bibfield  {author} {\bibinfo {author} {\bibfnamefont {M.}~\bibnamefont
  {Will}}, \bibinfo {author} {\bibfnamefont {M.}~\bibnamefont {Hamer}},
  \bibinfo {author} {\bibfnamefont {M.}~\bibnamefont {M{\"{u}}ller}}, \bibinfo
  {author} {\bibfnamefont {A.}~\bibnamefont {Noury}}, \bibinfo {author}
  {\bibfnamefont {P.}~\bibnamefont {Weber}}, \bibinfo {author} {\bibfnamefont
  {A.}~\bibnamefont {Bachtold}}, \bibinfo {author} {\bibfnamefont {R.~V.}\
  \bibnamefont {Gorbachev}}, \bibinfo {author} {\bibfnamefont {C.}~\bibnamefont
  {Stampfer}}, \ and\ \bibinfo {author} {\bibfnamefont {J.}~\bibnamefont
  {G{\"{u}}ttinger}},\ }\href {\doibase 10.1021/acs.nanolett.7b01845}
  {\bibfield  {journal} {\bibinfo  {journal} {Nano Letters}\ ,\ \bibinfo
  {pages} {5950}} (\bibinfo {year} {2017})}\BibitemShut {NoStop}%
\bibitem [{\citenamefont {Novoselov}\ \emph {et~al.}(2004)\citenamefont
  {Novoselov}, \citenamefont {Geim}, \citenamefont {Morozov}, \citenamefont
  {Jiang}, \citenamefont {Zhang}, \citenamefont {Dubonos}, \citenamefont
  {Grigorieva},\ and\ \citenamefont {Firsov}}]{NovoselovK01}%
  \BibitemOpen
  \bibfield  {author} {\bibinfo {author} {\bibfnamefont {K.~S.}\ \bibnamefont
  {Novoselov}}, \bibinfo {author} {\bibfnamefont {A.~K.}\ \bibnamefont {Geim}},
  \bibinfo {author} {\bibfnamefont {S.~V.}\ \bibnamefont {Morozov}}, \bibinfo
  {author} {\bibfnamefont {D.}~\bibnamefont {Jiang}}, \bibinfo {author}
  {\bibfnamefont {Y.}~\bibnamefont {Zhang}}, \bibinfo {author} {\bibfnamefont
  {S.~V.}\ \bibnamefont {Dubonos}}, \bibinfo {author} {\bibfnamefont {I.~V.}\
  \bibnamefont {Grigorieva}}, \ and\ \bibinfo {author} {\bibfnamefont {A.~A.}\
  \bibnamefont {Firsov}},\ }\href {\doibase 10.1126/science.1102896} {\bibfield
   {journal} {\bibinfo  {journal} {Science}\ }\textbf {\bibinfo {volume}
  {306}},\ \bibinfo {pages} {666} (\bibinfo {year} {2004})}\BibitemShut
  {NoStop}%
\bibitem [{\citenamefont {Blake}\ and\ \citenamefont {Hill}(2007)}]{BlakeP01}%
  \BibitemOpen
  \bibfield  {author} {\bibinfo {author} {\bibfnamefont {P.}~\bibnamefont
  {Blake}}\ and\ \bibinfo {author} {\bibfnamefont {E.~W.}\ \bibnamefont
  {Hill}},\ }\href {\doibase 10.1063/1.2768624} {\bibfield  {journal} {\bibinfo
   {journal} {Applied {P}hysics {L}etters}\ }\textbf {\bibinfo {volume} {91}},\
  \bibinfo {pages} {063124} (\bibinfo {year} {2007})}\BibitemShut {NoStop}%
\bibitem [{\citenamefont {Cui}\ and\ \citenamefont {Veres}(2008)}]{CuiB01}%
  \BibitemOpen
  \bibfield  {author} {\bibinfo {author} {\bibfnamefont {B.}~\bibnamefont
  {Cui}}\ and\ \bibinfo {author} {\bibfnamefont {T.}~\bibnamefont {Veres}},\
  }\href {\doibase 10.1016/j.mee.2008.01.008} {\bibfield  {journal} {\bibinfo
  {journal} {Microelectronic Engineering}\ }\textbf {\bibinfo {volume} {85}},\
  \bibinfo {pages} {810} (\bibinfo {year} {2008})}\BibitemShut {NoStop}%
\bibitem [{\citenamefont {{Castellanos-Gomez}}\ \emph
  {et~al.}(2014)\citenamefont {{Castellanos-Gomez}}, \citenamefont {Buscema},
  \citenamefont {Molenaar}, \citenamefont {Singh}, \citenamefont {Janssen},
  \citenamefont {{van der Zant}},\ and\ \citenamefont {Steele}}]{CGomez02}%
  \BibitemOpen
  \bibfield  {author} {\bibinfo {author} {\bibfnamefont {A.}~\bibnamefont
  {{Castellanos-Gomez}}}, \bibinfo {author} {\bibfnamefont {M.}~\bibnamefont
  {Buscema}}, \bibinfo {author} {\bibfnamefont {R.}~\bibnamefont {Molenaar}},
  \bibinfo {author} {\bibfnamefont {V.}~\bibnamefont {Singh}}, \bibinfo
  {author} {\bibfnamefont {L.}~\bibnamefont {Janssen}}, \bibinfo {author}
  {\bibfnamefont {H.~S.~J.}\ \bibnamefont {{van der Zant}}}, \ and\ \bibinfo
  {author} {\bibfnamefont {G.~A.}\ \bibnamefont {Steele}},\ }\href {\doibase
  10.1088/2053-1583/1/1/011002} {\bibfield  {journal} {\bibinfo  {journal} {2D
  Materials}\ }\textbf {\bibinfo {volume} {1}},\ \bibinfo {pages} {011002}
  (\bibinfo {year} {2014})}\BibitemShut {NoStop}%
\bibitem [{\citenamefont {{Teufel}}\ \emph {et~al.}(2011)\citenamefont
  {{Teufel}}, \citenamefont {{Donner}}, \citenamefont {{Li}}, \citenamefont
  {{Harlow}}, \citenamefont {{Allman}}, \citenamefont {{Sirois}}, \citenamefont
  {{Whittaker}}, \citenamefont {{Lehnert}},\ and\ \citenamefont
  {{Simmonds}}}]{Teufel1}%
  \BibitemOpen
  \bibfield  {author} {\bibinfo {author} {\bibfnamefont {J.~D.}\ \bibnamefont
  {{Teufel}}}, \bibinfo {author} {\bibfnamefont {T.}~\bibnamefont {{Donner}}},
  \bibinfo {author} {\bibfnamefont {D.}~\bibnamefont {{Li}}}, \bibinfo {author}
  {\bibfnamefont {J.~W.}\ \bibnamefont {{Harlow}}}, \bibinfo {author}
  {\bibfnamefont {M.~S.}\ \bibnamefont {{Allman}}}, \bibinfo {author}
  {\bibfnamefont {A.~J.}\ \bibnamefont {{Sirois}}}, \bibinfo {author}
  {\bibfnamefont {J.~D.}\ \bibnamefont {{Whittaker}}}, \bibinfo {author}
  {\bibfnamefont {K.~W.}\ \bibnamefont {{Lehnert}}}, \ and\ \bibinfo {author}
  {\bibfnamefont {R.~W.}\ \bibnamefont {{Simmonds}}},\ }\href {\doibase
  10.1038/nature10261} {\bibfield  {journal} {\bibinfo  {journal} {Nature}\
  }\textbf {\bibinfo {volume} {475}},\ \bibinfo {pages} {359 } (\bibinfo {year}
  {2011})}\BibitemShut {NoStop}%
\bibitem [{\citenamefont {L\'{e}v\^{e}que}\ and\ \citenamefont
  {{Villachon-Renard}}(1990)}]{LevequeG01}%
  \BibitemOpen
  \bibfield  {author} {\bibinfo {author} {\bibfnamefont {G.}~\bibnamefont
  {L\'{e}v\^{e}que}}\ and\ \bibinfo {author} {\bibfnamefont {Y.}~\bibnamefont
  {{Villachon-Renard}}},\ }\href {\doibase 10.1364/AO.29.003207} {\bibfield
  {journal} {\bibinfo  {journal} {Appl. Opt.}\ }\textbf {\bibinfo {volume}
  {29}},\ \bibinfo {pages} {3207} (\bibinfo {year} {1990})}\BibitemShut
  {NoStop}%
\bibitem [{\citenamefont {Katsidis}\ and\ \citenamefont
  {Siapkas}(2002)}]{KatsidisCC01}%
  \BibitemOpen
  \bibfield  {author} {\bibinfo {author} {\bibfnamefont {C.~C.}\ \bibnamefont
  {Katsidis}}\ and\ \bibinfo {author} {\bibfnamefont {D.~I.}\ \bibnamefont
  {Siapkas}},\ }\href {\doibase 10.1364/AO.41.003978} {\bibfield  {journal}
  {\bibinfo  {journal} {Appl. Opt.}\ }\textbf {\bibinfo {volume} {41}},\
  \bibinfo {pages} {3978} (\bibinfo {year} {2002})}\BibitemShut {NoStop}%
\bibitem [{\citenamefont {Byrnes}(2016)}]{ByrnesSJ01}%
  \BibitemOpen
  \bibfield  {author} {\bibinfo {author} {\bibfnamefont {S.~J.}\ \bibnamefont
  {Byrnes}},\ }\href {\doibase arXiv:1603.02720v2} {\bibfield  {journal}
  {\bibinfo  {journal} {arXiv}\ } (\bibinfo {year} {2016}),\
  arXiv:1603.02720v2}\BibitemShut {NoStop}%
\bibitem [{\citenamefont {{Raki{\'c}}}(1995)}]{RakicA01}%
  \BibitemOpen
  \bibfield  {author} {\bibinfo {author} {\bibfnamefont {A.~D.}\ \bibnamefont
  {{Raki{\'c}}}},\ }\href {\doibase 10.1364/AO.34.004755} {\bibfield  {journal}
  {\bibinfo  {journal} {Applied Optics}\ }\textbf {\bibinfo {volume} {34}},\
  \bibinfo {pages} {4755} (\bibinfo {year} {1995})}\BibitemShut {NoStop}%
\bibitem [{\citenamefont {Beal}\ \emph {et~al.}(1975)\citenamefont {Beal},
  \citenamefont {Hughes},\ and\ \citenamefont {Liang}}]{BealAR01}%
  \BibitemOpen
  \bibfield  {author} {\bibinfo {author} {\bibfnamefont {A.~R.}\ \bibnamefont
  {Beal}}, \bibinfo {author} {\bibfnamefont {H.~P.}\ \bibnamefont {Hughes}}, \
  and\ \bibinfo {author} {\bibfnamefont {W.~Y.}\ \bibnamefont {Liang}},\ }\href
  {\doibase 10.1088/0022-3719/8/24/015} {\bibfield  {journal} {\bibinfo
  {journal} {Journal of Physics {C}: Solid State Physics}\ }\textbf {\bibinfo
  {volume} {8}},\ \bibinfo {pages} {4236} (\bibinfo {year} {1975})}\BibitemShut
  {NoStop}%
\bibitem [{\citenamefont {Kness}(2012)}]{python_colorpy}%
  \BibitemOpen
  \bibfield  {author} {\bibinfo {author} {\bibfnamefont {M.}~\bibnamefont
  {Kness}},\ }\href@noop {} {\enquote {\bibinfo {title} {{colorpy 0.1.1}},}\
  }\bibinfo {howpublished} {https://pypi.python.org/pypi/colorpy} (\bibinfo
  {year} {27~Aug.~2012})\BibitemShut {NoStop}%
\bibitem [{\citenamefont {Sharma}\ \emph {et~al.}(2005)\citenamefont {Sharma},
  \citenamefont {Wu},\ and\ \citenamefont {Dalal}}]{SharmaG01}%
  \BibitemOpen
  \bibfield  {author} {\bibinfo {author} {\bibfnamefont {G.}~\bibnamefont
  {Sharma}}, \bibinfo {author} {\bibfnamefont {W.}~\bibnamefont {Wu}}, \ and\
  \bibinfo {author} {\bibfnamefont {E.~N.}\ \bibnamefont {Dalal}},\ }\href
  {\doibase 10.1002/col.20070} {\bibfield  {journal} {\bibinfo  {journal}
  {Color research and application}\ }\textbf {\bibinfo {volume} {30}},\
  \bibinfo {pages} {21} (\bibinfo {year} {2005})}\BibitemShut {NoStop}%
\bibitem [{\citenamefont {Jones}\ \emph {et~al.}(01  )\citenamefont {Jones},
  \citenamefont {Oliphant}, \citenamefont {Peterson} \emph {et~al.}}]{scipy}%
  \BibitemOpen
  \bibfield  {author} {\bibinfo {author} {\bibfnamefont {E.}~\bibnamefont
  {Jones}}, \bibinfo {author} {\bibfnamefont {T.}~\bibnamefont {Oliphant}},
  \bibinfo {author} {\bibfnamefont {P.}~\bibnamefont {Peterson}},  \emph
  {et~al.},\ }\href {http://www.scipy.org/} {\enquote {\bibinfo {title}
  {{SciPy}: Open source scientific tools for {Python}},}\ } (\bibinfo {year}
  {2001--}),\ \bibinfo {note} {[Online; accessed <today>]}\BibitemShut
  {NoStop}%
\bibitem [{\citenamefont {Malard}\ \emph {et~al.}(2009)\citenamefont {Malard},
  \citenamefont {Pimenta}, \citenamefont {Dresselhaus},\ and\ \citenamefont
  {Dresselhaus}}]{MalardLM01}%
  \BibitemOpen
  \bibfield  {author} {\bibinfo {author} {\bibfnamefont {L.}~\bibnamefont
  {Malard}}, \bibinfo {author} {\bibfnamefont {M.}~\bibnamefont {Pimenta}},
  \bibinfo {author} {\bibfnamefont {G.}~\bibnamefont {Dresselhaus}}, \ and\
  \bibinfo {author} {\bibfnamefont {M.}~\bibnamefont {Dresselhaus}},\ }\href
  {\doibase https://doi.org/10.1016/j.physrep.2009.02.003} {\bibfield
  {journal} {\bibinfo  {journal} {Physics Reports}\ }\textbf {\bibinfo {volume}
  {473}},\ \bibinfo {pages} {51 } (\bibinfo {year} {2009})}\BibitemShut
  {NoStop}%
\end{thebibliography}%

\end{document}